\documentclass[aps,prd,superscriptaddress,nofootinbib]{revtex4}

\usepackage{graphicx}

\usepackage{amsmath}

\usepackage{amssymb}

\usepackage{pstricks}

\begin{document}

\title{Crossing the cosmological constant line in a dilatonic brane-world model with and without curvature corrections}

\author{Mariam Bouhmadi-L\'{o}pez}
\email{mariam.bouhmadi@fisica.ist.utl.pt}
\affiliation{Centro Multidisciplinar de Astrof\'{\i}sica - CENTRA, Departamento de F\'{\i}sica, Instituto Superior T\'ecnico, Av. Rovisco Pais 1,
1049-001 Lisboa, Portugal}

\author{Antonio Ferrera}
\affiliation{Centro de F\'{\i}sica ``Miguel A. Catal\'{a}n'', Instituto de 
F\'{\i}sica Fundamental,
Consejo Superior de Investigaciones Cient\'{\i}ficas, Serrano 121, 28006 Madrid, Spain}

\begin{abstract}

We construct a new brane-world model composed of a bulk --with a
dilatonic field--, plus a brane --with brane tension coupled to the dilaton, cold dark matter and an induced gravity term. It is
possible to show that depending on the nature of the coupling between the brane tension and the dilaton this model can describe the late-time acceleration
of the brane expansion (for the normal branch) as it moves within the
bulk. The acceleration is produced 
together with a mimicry of the crossing of the cosmological constant
line ($w=-1$) on the brane, although this crossing of the phantom
divide is obtained without invoking any phantom matter neither on the
brane nor in the bulk. The role of dark energy is played by the brane
tension, which reaches a maximum positive value along the
cosmological expansion of the brane. It is precisely at that maximum
that the crossing of the phantom divide takes place. We also show that
these results remain valid when the induced gravity term on the brane
is switched off.
 
\end{abstract}

\date{\today}

\maketitle

\section{Introduction}\label{sec1}

This is a very exciting time for cosmology with the overwhelming amount of new observational data that theorists try to explain. One of the biggest puzzles that the theoretical community faces is to explain the recent speed up in the universe rate of expansion discovered  first through observations from distant type Ia supernova a decade ago \cite{Riess:1998cb,Perlmutter:1998np}. This late-time acceleration of the universe has been latter on confirmed by other independent observational probes  based for example on  measurement of  the cosmic microwave background radiation, the clustering of galaxies on very large scales and the baryon acoustic oscillations \cite{Tegmark:2003ud,Rapetti:2004aa,Spergel:2006hy,Percival:2007yw,Giannantonio:2008zi,Komatsu:2008hk}. 

A plethora of different theoretical models have been so far proposed to explain this phenomenon \cite{Copeland}, although unfortunately none of the models advanced so far is both completely convincing and well motivated. A cosmological constant corresponding to roughly two thirds of the total energy density of the universe is perhaps the simplest  way  to explain the late-time speed up of the universe --and in addition  would match rather well the observational data. However,  the expected theoretical value of the cosmological constant is about 120 orders of magnitude larger than the value needed to fit the data \cite{Durrer:2007re}. 

An alternative approach to explain the late-time acceleration is to invoke an infrared modification of general relativity  on large scales which, by weakening the gravitational interaction on those scales, allows the recent speed up of the universal expansion. This approach is also motivated by the fact that we only have precisions observations of gravity from sub-millimiter scales up to solar system scales while the Hubble radius, which is the scale relevant for the cosmic acceleration, is many orders of magnitude larger. 

A quite promising scheme in this approach is the Dvali, Gabadadze and Porrati (DGP) model \cite{Dvali:2000hr} which  corresponds to a 5-dimensional (5D) induced gravity brane-world model \cite{Deffayet,IG, Sahni:2002dx,LDGP2,Bouhmadi-Lopez:2004ys}, where a low-energy modification occurs with respect to general relativity; i.e. an infrared  effect takes place, leading to two  branches of solutions: (i) the self-accelerating branch and (ii)  the normal branch. 

As it name would suggest the self-accelerating branch solution gives
rise to a  late-time accelerating brane universe. The acceleration of
the brane expansion arises naturally, even without invoking the
presence of any dark energy on the brane to produce the speed-up. Not
surprisingly therefore this self-accelerated feature of the DGP model
has lead to considerable  research activity
\cite{reviewDGP}. Furthermore, it has been recently shown  that by
embedding the DGP model in a higher dimensional space-time the ghost
issue present in the original model \cite{Koyama:2007za} may be curred  \cite{deRham:2007xp} while preserving the existence of a self-accelerating solution \cite{Minamitsuji:2008fz}. 

The normal branch also constitutes in itself a very interesting result
of the DGP model however, as it can mimic a phantom-like behaviour on
the brane by means of the $\Lambda$DGP scenario
\cite{Sahni:2002dx}. At this respect we remind the reader that
observational data do not seem incompatible with a phantom-like
behaviour \cite{Percival:2007yw} and therefore we should keep an open
mind about what is producing the recent inflationary era of our
universe. On the other hand, this phantom-like behaviour may well be a
property acquired only recently by dark energy. This leads to an
interest in building models that exhibit the so called crossing of the
phantom divide line $w=-1$; for example in the context of the
brane-world scenario \cite{LDGP2,crossing}. The most important aspect
of the $\Lambda$DGP model is that the phantom-like mimicry is obtained
without invoking any real phantom-matter  \cite{phantom} which is
known to violate the null energy condition and induce quantum
instabilities\footnote{We are referring here to a phantom energy
  component described through a minimally coupled scalar field with
  the wrong kinetic term.} \cite{Cline:2003gs}. In the  DGP  scenario
it is also possible to obtain a mimicry of the crossing of the phantom divide at the cost of invoking a dynamical dark energy on the brane \cite{LDGP2}, for example modelled by a quiessence fluid or a (generalised)  Chaplygin gas.

In this paper we will show that there is an alternative form of mimicking the crossing the cosmological constant line $w=-1$ in the brane-world scenario. More precisely, we consider a 5D dilatonic bulk with a brane endowed with (or without)  an induced gravity term, a brane matter content corresponding   to cold dark matter, and a brane tension $\lambda$ that depends on the minimally coupled bulk scalar field. We will show that in this set-up the normal branch expands in an accelerated way due to $\lambda$ playing the role of \textit{dark energy} --through its dependence on the bulk scalar field. In addition, it turns out that $\lambda$ grows with the brane scale factor until it reaches a maximum positive value and then starts decreasing. Therefore, in our model the brane tension mimics crossing the phantom divide. Most importantly no matter violating  the null energy density is invoked in our model.

The paper can be outlined as follows: in section \ref{sec2} we
describe the general bulk plus brane scenario; in Section \ref{sec3}
we then analyse the vacuum (i.e., $\rho_{m}=0$) solution to
prepare the ground for the analysis carried out in
section \ref{sec4}. There we show that, under some assumptions on
the nature of the coupling parameters between $\lambda$ and $\phi$,
$1+w_{\rm{eff}}$ changes sign as the brane evolves, with $w_{\rm{eff}}$ the
effective equation of state for the brane tension. In section \ref{sec5} we show that the the presence of the
induced gravity term is not a necessary ingredient in order to mimic
the crossing of the phantom divide line  in this context. Our
conclusions are presented in section \ref{sec6}. Finally, in the
appendix \ref{appendix} we present an analytical  proof for the mimicry of the
crossing of the phantom divide in the model presented in section~\ref{sec2}.

\section{The setup}\label{sec2}

We consider a  brane, described by a 4D hyper-surface ($h$, metric g), embedded in a 5D bulk space-time ($\mathcal{B}$, metric 
$g^{(5)}$), whose action is given by 
\begin{eqnarray}
\mathcal{S} = \,\,\, \frac{1}{\kappa_5^2}\int_{\mathcal{B}} d^5X\, \sqrt{-g^{(5)}}\;
\left\{\frac{1}{2}R[g^{(5)}]\;+\;\mathcal{L}_5\right\}
 + \int_{h} d^4X\, \sqrt{-g}\; \left\{\frac{1}{\kappa_5^2} K\;+\;\mathcal{L}_4 \right\}\,, \label{action}
\end{eqnarray}
where $\kappa_5^2$ is the 5D gravitational constant,
$R[g^{(5)}]$ is the scalar curvature in the bulk and $K$ the extrinsic curvature of the brane in the higher dimensional
bulk, corresponding to the York-Gibbons-Hawking boundary term.  

We consider a dilaton  field $\phi$ living on the bulk \cite{Chamblin:1999ya,Maeda:2000wr} and we choose $\phi$ to be dimensionless. 
Then, the 5D Lagrangian $\mathcal{L}_5$ can be written as
\begin{eqnarray} 
\mathcal{L}_5=-\frac12 (\nabla\phi)^2 -V(\phi).
\end{eqnarray}

The 4D Lagrangian $\mathcal{L}_4$ corresponds to 
\begin{equation}
\mathcal{L}_4= \alpha {R}[g] -\lambda(\phi)+\Omega^{4}\mathcal{L}_m(\Omega^2 g_{\mu\nu}).
\label{L4}\end{equation}
The first term on the right hand side (rhs) of the previous equation corresponds to an induced gravity term \cite{Dvali:2000hr,Deffayet,IG, Sahni:2002dx},
where $R[g]$ is the scalar curvature of the induced metric on the
brane and $\alpha$ is  a positive parameter which  measures the strength of the induced gravity term 
and has dimensions of mass squared. The term $\mathcal{L}_m$ in Eq.~(\ref{L4}) describes  the matter content of the brane and $\lambda(\phi)$ is the brane tension, and we will restrict ourselves to the case where they are homogeneous and isotropic on the brane.
We allow the brane matter content to  be  non-minimally coupled on the (5D) Einstein frame  but to be minimally coupled respect to a conformal metric 
${\tilde g}^{(5)}_{AB}=\Omega^2\;g^{(5)}_{AB}$, where $\Omega=\Omega(\phi)$ \cite{Maeda:2000wr}. Notice that for reasons of mathematical simplicity we have not included a similar coupling to the induced gravity term.

We are mainly interested in the cosmology of this system. It is known that
for an expanding FLRW brane the unique bulk space-time in Einstein gravity (in vacuum) is a 5D Schwarzschild-anti de Sitter space-time 
\cite{BCG,MSM}. This property in principle cannot be extended  to a 5D dilatonic bulk.
On the other hand, we stress that the presence of an induced gravity term in the  brane-world scenario
affects only the dynamics of the brane, through the junction conditions at the brane, and does not affect the bulk field equations.
Consequently, in order to study the  effect of an induced gravity term in a brane-world dilatonic model, it is possible to consider a bulk
corresponding to a dilatonic 5D space-time and later on impose the junction conditions at the brane location. The junction 
conditions will then determine the dynamics of the brane and constrain the brane tension. This is the approach we will follow. 

From now on, we consider a 5D dilatonic solution obtained by Feinstein et al \cite{Feinstein:2001xs,Kunze:2001ji}
\textit{without an induced gravity term on the brane}. The 5D dilatonic solution reads \cite{Kunze:2001ji}
\begin{equation}
ds^2_5=\frac{1}{\xi^2}r^{2/3(k^2-3)}dr^2 +r^2(-d{t}^2+\gamma_{ij}dx^idx^j), 
\label{bulkmetric}\end{equation}
where  $\gamma_{ij}$ is a 3D spatially flat  metric. The bulk potential corresponds to
\begin{equation}
V(\phi)=\Lambda\exp[-(2/3) k\phi].
\label{liouville}\end{equation}
The parameters $k$ and $\xi$ in Eq.~(\ref{bulkmetric}) measure the magnitude of the 5D cosmological constant $\Lambda$ 
\begin{equation} \Lambda= \frac12 (k^{2}-12)\xi^2. \end{equation}
The 5D scalar field scales logarithmically with the radial coordinate $r$ \cite{Kunze:2001ji}
\begin{equation} \phi=k\log (r). \label{phi}\end{equation}

Now, we consider a FLRW brane filled only with cold dark matter (CDM); i.e pressureless matter,  and the brane tension $\lambda(\phi)$. As we will next show the late-time acceleration of the brane is driven by the brane tension through its dependence on the scalar field. On the other hand, the brane is considered to be embedded in the previous 5D dilatonic solution  
and its trajectory in the bulk is described by the following parametrisation 
\begin{equation}{t}={t}(\tau),\,\,\,\, r=a(\tau),\,\,\,\, x_i= constant,\,\, i=1 \ldots 3. \end{equation}
Here $\tau$ corresponds to the brane proper time. Then the brane metric reads
\begin{equation} 
ds^2_4\,=\,g_{\mu\nu}\,dx^{\mu}dx^{\nu}\,=\,-d\tau^2+a^2(\tau)\gamma_{ij}dx^idx^j.
\end{equation}
For an induced gravity brane-world model
\cite{Deffayet,Bouhmadi-Lopez:2004ys}, there are two physical ways of
embedding the brane in the bulk when a $\mathbb{Z}_2$-symmetry across
the brane is assumed. Here, we are interested in the cosmology of the
brane-world  whose geometry directly generalises that of the normal
DGP branch\footnote{We will refer to the normal DGP branch also as the non-self-accelerating DGP branch.}
because this geometrical construction allows a mimicry of the crossing of the cosmological constant line. Then, the location of the brane $r=a(\tau)$ is such that\footnote{Notice that the brane is moving in the bulk away from the bulk naked singularity located at $r=0$ \cite{Bouhmadi-Lopez:2004ys}.} 
\begin{equation}
ds^2_5=\frac{1}{\xi^2}r^{2/3(k^2-3)}dr^2 +r^2(-d{t}^2+\gamma_{ij}dx^idx^j),  \,\,\,\, r<a(\tau).\label{bulkmetric-}\end{equation}

For simplicity, we will consider that the matter content of the brane is minimally coupled respect to the  conformal metric ${\tilde g}^{(5)}_{AB}=\exp(2b\phi)\;g^{(5)}_{AB}$; i.e. $\Omega=\exp(b\phi)$, where $b$ is a constant.
We will also consider only the case\footnote{The main conclusions of the paper do not depend on the sign of $k$ but on the sign of the parameter $kb$. Therefore, we can always describe the same physical situation on the brane for $k<0$ by changing the sign of $b$.} $k > 0$; i.e. the scalar field is a growing function of the coordinate $r$.
Then, the Israel junction condition at the brane \cite{Chamblin:1999ya} describes the cosmological evolution of the brane through the modified Friedmann equation, which in our case reads
\begin{equation} \sqrt{{\xi^2}{a^{-\frac23 k^2}}+H^2} =\,\frac{\kappa_5^2}{6}
\left[\lambda(\phi)+\rho_m-6\alpha H^2\right].
\label{Friedmann1}\end{equation}
This equation will be crucial to prove the non super-acceleration of the brane; i.e. the Hubble rate decreases as the brane expands and moves in the bulk. The brane Friedmann equation can be more conveniently expressed as
\begin{eqnarray} 
H^2=
\frac{1}{6\alpha}\left\{\lambda+\rho_m
+\frac{3}{\kappa_5^4\alpha}
\left[1-\sqrt{1+4\kappa_5^4\alpha^2\xi^2a^{-2k^2/3}+\frac{2}{3}\kappa_5^4\alpha(\lambda+\rho_m)}\right]\right\},
\label{Friedmann}
\end{eqnarray}
where $\lambda$ is the brane tension and $\rho_m$ is the energy density of CDM.

On the other hand, as it is usual in a dilatonic brane-world scenario, matter on the brane --in this case CDM-- is not conserved due to the coupling $\Omega$ (see Eq.~(\ref{L4})). In fact, we have
\begin{equation}
\dot\rho_m=-3H(1-kb)\rho_m, 
\label{conservationrho}
\end{equation}
where a dot stands for derivatives respect to the brane cosmic time $\tau$. Therefore, CDM on the brane scales  as
\begin{equation}
\rho_m=\rho_0 a^{-3+kb}.
\label{dust}
\end{equation}

Finally, the junction condition of the scalar field at the brane \cite{Chamblin:1999ya} constrains the brane tension $\lambda(\phi)$. In our model this is given by
\begin{equation}
a\frac{d\lambda}{da}=-kb\rho_m\, {-}\frac{2k^2}{\kappa_5^2}\sqrt{{\xi^2}{a^{-\frac23 k^2}}+H^2},
\label{constraint}
\end{equation}
where for convenience we have rewritten the scalar field (valued at the brane) in terms of the scale factor of the brane. At this respect we remind the reader that  at the brane $\phi=k\log(a)$. 

Notice that for $kb>0$ the left hand side (lhs) term in Eq.~(\ref{constraint}) is always negative and therefore it always tends to decrease the brane tension. Therefore, a necessary condition for the brane tension to mimic a crossing of the phantom divide is that $kb<0$. This condition implies  that  CDM on the brane has to redshift faster that in the standard case, i.e. CDM on the brane has to decay faster than $a^{-3}$ (see Eq.~(\ref{dust})). In order to see what happens in this case it is useful to analyse first the vacuum case; i.e. when $\rho_m=0$.

\section{Vacuum solution}\label{sec3}

For a vacuum brane; i.e. for $\rho_m=0$, the brane tension is a decreasing function of the scalar field or equivalently of the scale factor of the brane\footnote{The constraint equation (\ref{constraint}) (after substituting the Hubble rate given in Eq.~(\ref{Friedmann})) can be solved analytically in this case \cite{Bouhmadi-Lopez:2004ys} and it can be explicitly shown that the brane tension decreases as the brane expands. In  the same way a parametric  expression can be found for the Hubble rate and its cosmic derivative. \label{footnote3}} (see Eq.~(\ref{constraint})). For small values of the scale factor, the brane tension reaches infinite positive values. On the other hand, for very large value of the scale factor the brane tension vanishes. 

The Hubble parameter is  a decreasing function of the scale factor, i.e the brane is never super-accelerating. In fact,
\begin{equation}
\dot H=-\frac{ k^2 H^2}{\kappa_5^4\alpha(\lambda-6\alpha H^2)+3},
\end{equation}
while the Israel junction condition (\ref{Friedmann1}) implies that the denominator of the previous equation has to be positive (see also footnote \ref{footnote3}), therefore $\dot H<0$. At high energy, $H$ reaches  a constant positive value. Consequently, in the vacuum brane there is no big bang singularity on the brane; indeed, the brane is asymptotically de Sitter. On the other hand, at very large values of the scale factor, the Hubble parameter vanishes (the brane is asymptotically Minkowski in the future). Although the brane never super-accelerates, the brane always undergoes an inflationary period. 

The brane behaves in two different ways depending on the value taken
by $k^2$. Thus, for $k^2\leq 3$ the brane is eternally inflating. A
similar behaviour was found in \cite{Kunze:2001ji}.  On the other
hand, for $k^2>3$ the brane undergoes an initial stage of inflation
and later on it starts decelerating. This second behaviour contrasts
with the results in \cite{Kunze:2001ji} for a vacuum brane without an
induced gravity term on the brane. Then, the inclusion of an induced gravity term on a dilatonic brane-world model with an exponential potential in the bulk allows for the normal branch to inflate  in a region of parameter space where the vacuum dilatonic brane alone would not inflate. This behaviour has some similarity with steep inflation \cite{Copeland:2000hn}, where high energy corrections to the Friedmann equation in RS scenario \cite{Randall:1999vf,reviewRS} permit an inflationary evolution of the brane with potentials too steep to sustain it in the standard 4D case, although the inflationary scenario introduced by Copeland et al in  \cite{Copeland:2000hn} is supported by an inflaton confined in the brane while in our model inflation on the brane is induced by a dilaton field on the bulk. 

The important thing we have learned from the vacuum solution for our next step of model building is that the brane is eternally inflating if $k^2\leq 3$. We will next show that if this condition holds in a brane filled with CDM (together with others conditions that we will next enumerate), the brane tension will mimic a crossing of the cosmological constant line.

\section{Crossing the phantom divide}\label{sec4}

We now address the main point of the paper: is it possible to mimic a crossing of the phantom divide in particular in the model introduced in section \ref{sec2}? Unlike the vacuum case --which can be solved analytically \cite{Bouhmadi-Lopez:2004ys}-- in this case we cannot exactly solve the constraint (\ref{constraint}). Nevertheless, we can answer the previous question based in some physical and reasonable assumptions  and as well as on numerical methods. For technical details of the analytical proof we refer the reader to the appendix section. These are assumptions that we make:

\begin{enumerate}
\item We assume that CDM dominates over the vacuum term ($a^{-2/3
    k^2}$) at early times on the brane. This implies that the
  parameter $\beta_0$ introduced in Eq.~(\ref{def}) (see the Appendix) must satisfy $\beta_0 > 1$. On the other hand, the brane tension will play the role of dark energy (through its dependence on the scalar field) in our model. This first assumption assumes that dark matter dominates over dark energy at high redshift which is a natural assumption to make.

\item We also assume that CDM redshifts away a bit faster than usual; i.e. $bk<0$ or $\beta_2$ introduced in Eq.~(\ref{def}) is such that $\beta_2 > 1$. This lost energy will be used to increase the value of the scalar field $\phi(a)$ on the brane. That is, to push the brane to higher values of $a$. 
\item Finally, we also assume that $\beta_2< 2\beta_0(\beta_2-1)$. This condition, together with $\beta_0, \beta_2 >  1$, is sufficient to prove the non existence of a local minimum of the brane tension during the cosmological evolution of the brane. In fact, as we show in the appendix the existence of a unique maximum can be proven for an even larger set of parameters $\beta_0>1/2$,  $\beta_2 > 1$ and $\beta_2< 2\beta_0(\beta_2-1)$. Therefore,
the set of allowed parameter $k$ and $b$ that fulfil the last three inequalities
are such that
\begin{equation}
k< {\rm min}\left\{-3b,\frac32\left[-b+\sqrt{b^2+4}\right]\right\}=-3b.
\end{equation}
where $b$ is positive.
\end{enumerate}
 
\begin{figure}[h]
\begin{center}
\includegraphics[width=6cm]{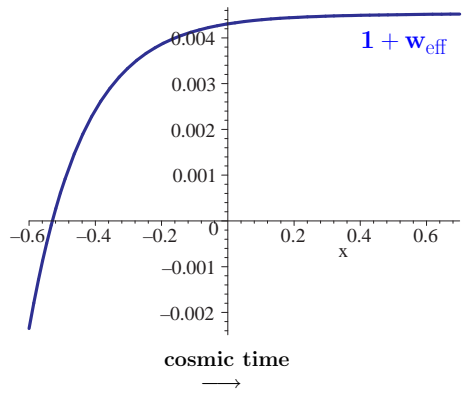}
\end{center}
\caption{The figure shows the effective equation of state of the brane tension defined in Eq.~(\ref{eqstate}) against the variable $x$ defined in Eq.~(\ref{def}). Notice that $x$ grows as the brane expands and therefore $dx/d\tau>0$ where $\tau$ corresponds to the cosmic time of the brane. This illustrative numerical solution has been obtained for $b=-1$, $k=1$ and $\beta_1=1$. The last parameter is defined in Eq.~(\ref{def}). In order to impose the right initial condition, we started the integration well in the past where CDM dominated over the scalar field on the brane and we took as a good approximated solution the dark matter solution given in Eq.~(\ref{asymlambda11}).}
\label{efectomenweff}
\end{figure}

Under these three assumptions, it can be shown that the brane tension has a local maximum which must be positive (we refer the reader again to the appendix section for this proof). In fact, what happens under these conditions is that the brane tension increases until it reaches its maximum positive value and then it starts decreasing. It is precisely at this maximum that the brane tension mimics crossing the phantom divide. Around the local maximum of the brane tension we can always define an effective equation of state in analogy with the standard 4D relativistic case:
\begin{equation}
1+w_{\textrm{eff}}=-\frac{1}{3H}\frac{1}{\lambda}\frac{d\lambda}{d\tau}.
\label{eqstate}
\end{equation}
As we mentioned earlier, the constraint equation (\ref{constraint}) cannot be solved analytically and therefore we have to resort to numerical methods. We show in figure \ref{efectomenweff} an example of  our numerical results where it can be seen clearly that $1+w_{\textrm{eff}}$ changes sign. It is precisely at that moment that the crossing takes place.

\begin{figure}[h]
\begin{center}
\includegraphics[width=6cm]{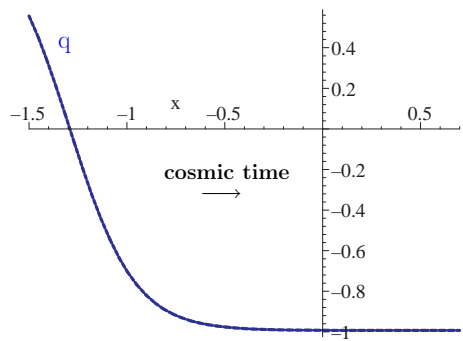}
\end{center}
\caption{The figure shows the deceleration  parameter $q=-\ddot a a/\dot a^2$ against the variable $x$ defined in Eq.~(\ref{def}). The brane is accelerating in the future when $q$ is negative. Notice that $x$ grows as the brane expands and therefore $dx/d\tau>0$ where $\tau$ corresponds to the cosmic time of the brane. This numerical example has been obtained for $b=-1$, $k=1$ and $\beta_1=1$. The last parameter is defined in Eq.~(\ref{def}). Again in order to impose the right initial condition, we started the integration well in the past where CDM dominated over the scalar field on the brane and we can take  as a good approximated solution the dark matter solution given in Eq.~(\ref{asymlambda11}).}
\label{efectomenq}
\end{figure}

Another important question to address is whether the brane is accelerating at the time that the crossing takes place. We know that the vacuum term dominates at late times (see the first assumption). Thus, at that point the brane tension will be adequately described by the vacuum solution. From the results in the previous section then we can conclude that the brane will be speeding up at late times  as long as  $k^2\leq3$. On the other, hand it can be checked numerically that the brane can be accelerating at the crossing as the figure \ref{efectomenq} shows.

\section{Is induced gravity a necessary ingredient in this sort of model building?}\label{sec5}

We analyse next the same physical situation of section \ref{sec4} but now assuming that $\alpha=0$; i.e. we assume the dilatonic brane-world model introduced in Ref.~\cite{Kunze:2001ji} where matter corresponds to CDM. Unlike a vacuum brane  and a radiation filled brane, the CDM filled brane was not analysed in details in Ref.~\cite{Kunze:2001ji}. 

The Friedmann equation can be retrieved by setting $\alpha\rightarrow 0$ in Eq.~(\ref{Friedmann}). This limit is well defined for the normal branch that we have analysed previously. Then, the Hubble rate decreases as the brane expands and moves in the bulk such that 
\begin{equation}
H^2=\frac{\kappa_5^4}{36}\left(\lambda(\phi)+\rho_m\right)^2 -\xi^2a^{-\frac23 k^2},
\label{Friedmanna0}
\end{equation}
where CDM redshift as in Eq.~(\ref{dust}). The reason why CDM (and any matter in general) red-shifts as in the case of induced gravity is because we have assumed in section \ref{sec2} that the induced gravity term in the brane action is not coupled to the bulk scalar field, i.e. $\alpha$ is constant.

Now, by using Eqs.~(\ref{Friedmann1}) and (\ref{constraint}) and setting $\alpha\rightarrow 0$, we recover the constraint that must satisfy the brane tension in this particular case \cite{Kunze:2001ji}
\begin{equation}
a\frac{d\lambda}{da}=-kb\rho_m\, {-}\frac{k^2}{3}(\lambda+\rho_m).
\label{constraintwithoutig}
\end{equation}
The Israel junction condition imposes that $\lambda+\rho_m$ has to be positive\footnote{This is related to the way of embedding the brane in the bulk. In this particular case, if we use Gaussian normal coordinate, it turns out that the warping of the geometry is such that the warp factor acquires it maximum at the location of the brane. In this sense, we would have a sort of geometry similar to the Randall and Sundrum (RS) model \cite{Randall:1999vf}. We remind the reader that the brane tension is positive in RS model and in particular for a vacuum brane; i.e. for $\rho=0$.}, c.f. Eq.~(\ref{Friedmann1}) with $\alpha=0$. Therefore, if $kb>0$ the constraint (\ref{constraintwithoutig}) implies that the brane tension decreases as the brane expands. So, a necessary condition for the brane tension to mimic  a crossing of the phantom divide is  $kb<0$. This  is the  same condition we reached in section \ref{sec2} for  the induced gravity setup. Having $kb<0$ corresponds to the fact that matter on the brane redshift faster than in the standard relativistic case.

Notice that if the brane tension constraint is written in terms of
energy densities defined on the brane; i.e $\lambda$, $\rho_m$ and the
vacuum energy density proportional to $a^{-\frac23 k^2}$, then the
constraint for the induced gravity case involves the vacuum energy
density (cf. Eq.~(\ref{constraint})) while in absence of the induced
gravity term on the brane this energy density is absent in the
constraint equation (cf. Eq.~(\ref{constraintwithoutig})). This
feature implies that 
the constraint given in Eq.~(\ref{constraintwithoutig}) can be solved analytically  \cite{Kunze:2001ji}
\begin{equation}
\lambda(a)=\lambda_0 a^{-\frac{k^2}{3}}-\left(\frac{k(k+3b)}{k^2+3kb-9}\right)\rho_0 a^{-3+kb}; \qquad \lambda_0,\rho_0={positive\,\,\,\ constant} .
\label{constraintwithoutig2}
\end{equation}
Now, it is much easier to impose the three assumptions assumed in the previous sections:

\begin{enumerate}

\item We assume that the second term on the rhs of Eq.~(\ref{constraintwithoutig2}) dominates over the first term at early time, i.e. the early time evolution of the brane tension is driven by the CDM confined on the brane while the late-time evolution of the brane is driven by the vacuum term. This translates also into the fact that the early time evolution is CDM dominated while it is the vacuum who drives the late-time evolution [cf. Eq.~(\ref{Friedmanna0})]. This condition translates into a constraint in $\beta_0$, introduced in Eq.~(\ref{def}), $\beta_0>\frac12$ which implies
\begin{equation}
k^2<9-3kb.
\label{kmin1}
\end{equation}
Our first assumption also implies
\begin{equation}
\kappa_5^4\lambda_0^2-36\xi^2>0,
\end{equation}
so that the square of the Hubble parameter is well defined  at late-time (large $a$). On the other hand, our first assumption also implies that the Hubble parameter cannot vanishes at any finite value of the radius of the brane; i.e. the whole evolution of the brane is Lorentzian.
\item We also assume that CDM redshifts away a bit faster than usual; i.e. $bk<0$ or $\beta_2$ introduced in Eq.~(\ref{def}) is such that $\beta_2 > 1$.

\item Finally, we assume that the brane tension is initially negative. This last condition, Eqs.~(\ref{constraintwithoutig2}) 
and (\ref{kmin1}) imply
\begin{equation}
k<-3b.
\label{kmin2}
\end{equation}
Even though the brane tension is initially negative the sum of the CDM energy density plus the brane tension is always positive.  Therefore, the Israel junction condition Eq.~(\ref{Friedmann1}) with $\alpha=0$ is never violated.
\end{enumerate}
In summary, under the three conditions enumerated above we have that for a  given parameter $b$ the set of allowed values of $k$ are constrained to satisfy (cf. Eqs.~(\ref{kmin1}), (\ref{kmin2}) and $kb<0$)
\begin{equation}
0<k<{\rm{min}}\left\{-3b,\,\frac32\left[-b+\sqrt{b^2+4}\right]\right\}=-3b,
\label{kset}
\end{equation}
more importantly, this condition implies that the brane tension mimics a crossing of the phantom divide  as we next show.
We remind that an effective equation of state related to the brane tension can be defined as in Eq.~(\ref{eqstate}) or   equivalently as 
\begin{equation}
a\frac{d\lambda}{da}=-3(1+w_{\rm{eff}})\lambda.
\label{eqstate2}
\end{equation}
The crossing of the cosmological constant line happens when the rhs of Eq.~(\ref{eqstate2})  vanishes and $\lambda\neq 0$.  By combining the constraint satisfied by the brane tension (\ref{constraintwithoutig}) and Eq.~(\ref{eqstate2}), such a crossing takes place at the scale factor $a_{\rm{cross}}$ where 
\begin{equation}
a_{\rm{cross}}=\left[\frac{\lambda_0}{\rho_0}\frac{k(k^2+3kb-9)}{3(3-kb)(k+3b)}\right]^{\frac{3}{k^2-9 +3kb}}.
\label{across}
\end{equation}
As can be checked this scale factor is well defined if the parameter $k$ --related to the slope of the bulk scalar field-- is within the range (\ref{kset}).  On the other hand, the brane tension is positive at the crossing because (see for example Eqs.~(\ref{constraintwithoutig}) and (\ref{eqstate2})) at such event
\begin{equation}
\lambda=-\frac{1}{k}\left(3b+k\right)\rho_m,
\end{equation}
and by virtue of the previous three assumptions  the rhs term of the last equation is positive and therefore the brane tension at the crossing. This means that ever since the brane tension becomes positive, $\lambda$ has a phantom-like behaviour.Then, for scale factors larger than $a_{\rm{cross}}$ --where the crossing takes place--   $\lambda$ has a quintessence-like behaviour. The larger is the amount of CDM; i.e. the larger is $\rho_0$, the crossing takes place for a larger scale factor [cf. Eq.~(\ref{across})]. The effect of $\lambda_0$ is the opposite, i.e. the larger is $\lambda_0$ then the crossing takes place at a smaller scale factor. 

In summary, it is possible to have a crossing of the phantom divide in this sort of brane-world models without an induced gravity term on the brane action. 
It remains to be checked if the brane is accelerating when the crossing takes place with model parameters that are consistent with the constraints coming from the big bang nucleosynthesis  \cite{Kunze:2001ji} and the CDM equation of state \cite{Muller:2004yb}. 

The second derivative of the scale factor respect to the cosmic time reads
\begin{eqnarray}
\ddot a= A_0 a^{1-\frac23 k^2}+A_1a^{-2-\frac{k^2}{3}+kb}+A_2 a^{-5+2kb}
\label{aceleracion}
\end{eqnarray}
where
\begin{eqnarray}
  A_0=\left(\frac{\kappa_5^4\lambda_0^2}{36}-\xi^2\right)\left(1-\frac13 k^2\right), \,\,\, A_1=\frac14
\kappa_5^4\lambda_0\rho_0\frac{1-kb+\frac{k^2}{3}}{k^2+3kb-9}, \,\,\, A_2=\frac94\kappa_5^4\rho_0^2 \frac{-2+kb}{(k^2+3kb-9)^2}.
\end{eqnarray}

At large scale factors and  under the assumptions we have made, the second derivative of the scale factor is dominated by the first term on the rhs of Eq.~(\ref{aceleracion}). Therefore, the brane is accelerating for large scale factor as long as $k^2<3$. Of course, this  is not surprising and it is a consequence of the fact that the brane is vacuum dominated at late-time and for a vacuum brane it was shown in \cite{Kunze:2001ji} that the brane is eternally inflating if $k^2<3$. On the other hand, at early time  the expansion of the brane is dominated by the last term  on the rhs; i.e. by CDM on the brane, therefore the brane is decelerating initially. So, in summary the brane transits from a decelerating regime at early-time to an accelerating regime at late-time for $k^2<3$. On the other hand, an observational  consistency of these sort of brane-world models since the epoch of nucleosynthesis requires that  the quadratic term on the energy density is negligible respect to the linear term on the energy density (since the time of  nucleosynthesis)   \cite{Kunze:2001ji}  (for a review see for example \cite{reviewRS}).  We will assume this is the case and therefore the last term on Eq.~(\ref{aceleracion})  is negligible respect to the second one on the rhs. Consequently, the brane will start accelerating from a scale factor 
$a_{\textrm{acc}}$ onward where
\begin{equation}
a_{\textrm{acc}}\simeq \left|\frac{A_1}{A_0}\right|^{\frac{3}{9-3kb-k^2}}.
\label{acc}
\end{equation}
Therefore, as long as $a_{\textrm{acc}}$ is of the order of $a_{\rm{cross}}$, the starting speed up of the brane and the crossing of the phantom divide by the brane tension will be more or less at the same epoch.
In practice, this will require a fine tuning of the parameters of the model as we next show.

We will consider that the current scale factor is set to one. Then, by combing the following observational constraints:

\begin{enumerate}

\item The equation of state of dark matter is constrained in such a way that it does not deviate too much from the equation of state of a pressureless fluid in the standard 4D relativistic case (for recent bounds on the equation of state of dark matter see Ref.~\cite{Muller:2004yb}). As an estimate we consider the (strongest) constraint obtained in  Ref.~\cite{Muller:2004yb} which implies that
\begin{equation}
|kb|< 10^{-6}.
\end{equation}

\item The quadratic term on the energy density is negligible since the epoch of nucleosynthesis \cite{Kunze:2001ji}  

\begin{equation}
\frac{\rho_0}{\lambda_0}\ll 2(1-\frac13 kb -k^2)a_{\rm{nucl}}^{3-kb-\frac13 k^2}
\end{equation}

\item In addition, as it is usual in dilatonic models the effective gravitational \textit{constant} on the brane is not constant. This also implies a constraint on the parameter $k$  such that the effective gravitational \textit{constant} has not varied too much since the epoque of nucleosynthesis\footnote{We take $h=0.72$ from the latest WMAP result\cite{Komatsu:2008hk}.} \cite{Kunze:2001ji} 

\begin{equation}
k^2\leq 0.04.
\end{equation}

\item The amount of dark energy is about 3 times the amount of CDM \cite{Spergel:2006hy}; i.e. $\Omega_{de}\simeq 0.75$ and 
\begin{equation}
\frac{\Omega_{\rm{de}}}{\Omega_{\rm{cdm}}}=\left({1-\frac13kb-\frac19k^2}\right)\frac{\kappa_5^4\lambda_0^2-36\xi^2}{2\kappa_5^4\lambda_0\rho_0}\simeq 3.
\end{equation}

\end{enumerate}

Therefore, the constraints 1 and 3   implies that $|bk|$ and $k$ have to be very small. This fact combined with the conditions 2 and 4 implies:
\begin{equation}
\frac{\rho_0}{\lambda_0}\ll 10^{-30}, \,\,\,\, {\kappa_5^4\lambda_0^2-36\xi^2}\simeq 6 {\kappa_5^4\lambda_0\rho_0},
\label{constraintlr}
\end{equation}
where we have roughly estimated the size (radius) of the Universe at the time of nucleasinthesis, $a_{\rm{nucl}}$, to be  $10^{-10}$ smaller than today. Then we can conclude that the brane starts accelerating since it has half of its present size (cf. Eq.~(\ref{acc}))
\begin{equation}
a_{\textrm{acc}}\sim 0.55, 
\end{equation}
or at a redshift $z_{\textrm{acc}}\sim 0.8$. On the other hand, the constraint on the ratio $\rho_0/\lambda_0$ (see Eq.(\ref{constraintlr})) translates into a constraint 
\begin{equation}
  \left|\frac{k}{b}\right|\simeq \frac{\rho_0}{\lambda_0}\ll 10^{-30},
\end{equation}
for the crossing of the phantom divides by the brane tension to take place roughly at the same time when the brane starts speeding up. This strong fine tuning is inherited from imposing that the model is consistent since the epoque of nucleosynthesis. Notice that (i) if  this fine tuning is relaxed then the crossing of the phantom divide would take place in the past  of the brane; i.e the brane tension would been currently decreasing  as the brane expands and would have  a quintessence behaviour at present and (ii) if the fine tuning is made stronger then the brane tension would be still growing as the brane expands and the crossing of the phantom divide would take place in the future evolution of the brane.

\section{Conclusions}\label{sec6}

In this paper we have shown the existence of a mechanism that mimics the crossing of the cosmological constant line $w=-1$ in the brane-world scenario, and which is different from the one introduced in Refs.~\cite{Sahni:2002dx,LDGP2}. More precisely, we have shown that if we have a 5D dilatonic bulk with or without an induced gravity term on the brane (normal branch), a brane tension $\lambda$ which depends on the minimally coupled bulk scalar field, and a  brane matter content corresponding only to cold dark matter, then under certain conditions the brane tension grows with the brane scale factor until it reaches a maximum positive value at which it mimics crossing the phantom divide, and then starts decreasing. Most importantly no matter violating  the null energy condition is invoked in our model. Despite of the transitory phantom-like behaviour of the brane tension no big rip singularity is hited along the brane evolution.

In the model with an induced gravity term on the brane (normal branch or non-self-accelerating branch), the constraint equation fulfilled by the brane tension is too complicated to be solved analytically (see Eqs.~(\ref{Friedmann}) and (\ref{constraint})). However, we have shown that under certain physical and mathematical conditions -cold dark matter dominates at higher redshifts and it dilutes a bite faster than dust during the brane expansion as well as a mathematical condition that guarantees the non-existence of a local minimum of the brane tension- it is possible for the brane tension to cross the cosmological constant line. The analytical proof (see the appendix) has been confirmed by numerical solutions.  Furthermore, we have shown that for some values of the parameters the normal branch inflates eternally to the future due to the brane tension $\lambda$ playing the role of dark energy through its dependence on the bulk scalar field.

On the other hand, in the model without an induced gravity term things are much easier to analyse as it is possible to get an analytical solution for the brane tension \cite{Kunze:2001ji}. For the analytical solution and under the same physical and mathematical conditions assumed in the model with an induced gravity effect, we have shown that the brane tension grows until it reaches its maximum positive value and then it starts decreasing driving the late-time acceleration of the brane. It is precisely at that maximum value that the brane tension mimics the crossing of the phantom divide as its effective equation of state parameter $w_{\rm{eff}}$ is such that $1+w_{\rm{eff}}$ changes its sign in a smooth way.
We have also imposed observational bounds -from the dark matter equation of state  \cite{Muller:2004yb} and  the big bang nucleosynthesis \cite{Kunze:2001ji} that constraints the modified Friedmann equation and the effective gravitational constant of the brane- to constraint the parameters of the model.

In summary, in the model presented here the mimicry of the phantom
divide crossing is based on the interaction between the brane and the
bulk through the brane tension that depends explicitly on the scalar
field that lives in the bulk. We have also shown that the brane
undergoes a late-time acceleration epoch.

\section*{Acknowledgements}

The authors are grateful to  K.~E.~Kunze and M.~A.~V\'azquez-Mozo for helpful comments.
MBL is  supported by the Portuguese Agency Funda\c{c}\~{a}o para a Ci\^{e}ncia e
Tecnologia through the fellowship SFRH/BPD/26542/2006. 

\appendix 

\section{Proof of the existence of a mimicry of the crossing in the
  induced gravity  scenario}\label{appendix}

In this appendix we prove the existence of a mimicry of crossing the phantom divide modelled through the brane tension. For the proof it is useful to introduce the following dimensionless quantities
\begin{eqnarray}
&&\bar\lambda\equiv\frac23\kappa_5^4\alpha\lambda,\,\, x\equiv \frac23 k\phi-\ln d, \,\, d\equiv 4\alpha^2\kappa_5^4 \xi^2,\,\, m\equiv 3-kb,\nonumber\\
&&\beta_0\equiv\frac{9\beta_2}{2k^2},\,\,\beta_1\equiv\frac{2\kappa_5^4\alpha}{m}\rho_0d^{-\beta_0},\,\, \beta_2\equiv\frac{m}{3}. \label{def}
\end{eqnarray}
In terms of these new variables, the constraint given in Eq.~(\ref{constraint}) reads
\begin{eqnarray}
\frac{d\bar\lambda}{dx}=1-\beta_0\beta_1(1-\beta_2)e^{-\beta_0x}-
\sqrt{1+\bar\lambda+e^{-x}+\beta_1\beta_2 e^{-\beta_0x}}.
\label{dvdx}
\end{eqnarray}

If $\beta_0>1$; i.e. if  we assume that the CDM energy density (proportional to $\exp(-\beta_0x)$) dominates at early time over the vacuum term (the term $\exp(-x)$ is due to the bulk scalar field) then it can be shown that the brane tension
scales at that epoch as 
\begin{equation}
\bar\lambda\sim\beta_1(1-\beta_2)e^{-\beta_0 x} + O(e^{-\beta_0 x/2 }).
\label{asymlambda11}
\end{equation}
The same asymptotic behaviour is found for $1/2<\beta_0<1$ 
\begin{equation}
\bar\lambda\sim\beta_1(1-\beta_2)e^{-\beta_0 x} + O(e^{-x/2}),
\label{asymlambda12}
\end{equation}
being the subleading term different.

Furthermore, if we assume in addition that the parameter $\beta_2>1$; i.e. CDM redshifts faster than in the 4D standard relativistic case, then the brane tension is negative at early times and it is an increasing function of the scale factor (cf. Eqs.(\ref{asymlambda11}) and (\ref{asymlambda12})). 

On the other hand, the condition $\beta_0>1$ implies also that the
vacuum term due to the scalar field will dominate over the CDM term
at late-time and it can be proven that at large values of $x$ (or equivalently large values of the scale factor) 
\begin{equation}
\bar\lambda\sim C\exp(-x/2) + \ldots\,,\quad C=constant\,>0.
\label{asymlambda21}
\end{equation}
The constant $C$ is positive because for the vacuum solution the brane tension is always positive \cite{Bouhmadi-Lopez:2004ys}. Again, if $1/2<\beta_0<1$ it can be proven that the asymptotic behaviour of the dimensionless tension $\bar\lambda$ for $1\ll x$ is the same as for $1<\beta_0$
given in Eq.~(\ref{asymlambda21}). Therefore, at very late-time the brane tension decreases as the brane expands (see Eq.~(\ref{asymlambda21})). 

Now, since (i) $\bar\lambda$  is negative and an increasing function of $x$ at $\tau\ll 1$,  and (ii) $\bar\lambda$ is a positive valued, decreasing function of $x$ at $\tau\gg 1$, then the brane tension must have at  least a local positive maximum (we remind the reader that the brane tension is a smooth  analytical function of $x$). It is precisely at this maximum that the mimicry of crossing the phantom divide takes place.

One can go a step further and ask the following question: is there a local minimum such that $\beta_0, \beta_2 > 1$, and $\beta_2< 2\beta_0(\beta_2-1)$? The answer is no. This can be proven using a {\it{Reductio ad absurdum}}. We start assuming the existence of a local minimum of $\bar\lambda$  at $x_0$ then at the minimum
\begin{equation}
\frac{d^2\bar\lambda}{dx^2}|_{x_0}=\frac{e^{-x_0}-2\beta_0^3\beta_1^2(1-\beta_2)^2 e^{-2\beta_0 x_0}
+\beta_0\beta_1\left[\beta_2+2\beta_0(1-\beta_2)\right]e^{-\beta_0x_0}}{2\left[1-\beta_0\beta_1(1-\beta_2)e^{-\beta_0x_0}\right]}
\label{2deriv}
\end{equation}
The denominator of the previous equation is positive because $\beta_2>1$. On the other hand, (i) as the expression
\begin{equation}
\beta_0\beta_1\left[\beta_2+2\beta_0(1-\beta_2)\right]e^{-\beta_0x_0}<0,
\end{equation}
because we have assumed  $\beta_2< 2\beta_0(\beta_2-1)<0$, and (ii) the expression
\begin{equation}
e^{-x_0}-2\beta_0^3\beta_1^2(1-\beta_2)^2 e^{-2\beta_0 x_0}<0
\end{equation}
because the square of the Hubble parameter is positive at $x_0$ and $1/2<\beta_0$, then
it turns out that the second derivative given in Eq.~(\ref{2deriv}) is negative; i.e. at $x_0$ the dimensionless 
brane tension has no local minimum.

Therefore, under these conditions; $1/2<\beta_0$, $1<\beta_2$ and  $\beta_2\leq 2\beta_0(\beta_2-1)$, we know that the local maximum whose existence we have just proven is the unique extremum of $\bar\lambda$. This fact is confirmed by our numerical results.

\end{document}